\documentclass{mem}
\usepackage{natbib}
\usepackage{txfonts}
\usepackage{balance}
\usepackage{graphicx}
\usepackage[a4paper]{hyperref}

\begin{document}

\title{An Asteroseismic Test of Diffusion Theory}

\author{Travis S.~Metcalfe}

\institute{High Altitude Observatory, NCAR, P.O. Box 3000, Boulder CO 
80307-3000 USA}

\authorrunning{Metcalfe}
\titlerunning{Testing Diffusion Theory}

\abstract{The helium-atmosphere (DB) white dwarfs are commonly thought to 
be the descendants of the hotter PG~1159 stars, which initially have 
uniform He/C/O atmospheres. In this evolutionary scenario, diffusion 
builds a pure He surface layer which gradually thickens as the star cools. 
In the temperature range of the pulsating DB white dwarfs ($T_{\rm 
eff}\sim 25,000$~K) this transformation is still taking place, allowing 
asteroseismic tests of the theory. Objective global fitting of our updated 
double-layered envelope models to recent observations of the pulsating DB 
star CBS~114, and to existing observations of the slightly cooler star 
GD~358, lead to determinations of the envelope masses and pure He surface 
layers that qualitatively agree with the expectations of diffusion theory. 
These results provide new asteroseismic evidence supporting one of the 
central assumptions of spectral evolution theory, linking the DB white 
dwarfs to PG~1159 stars.

}

\maketitle{}

\section{Introduction}

In January 2002, the star V838~Mon suddenly became 600,000 times more 
luminous than our Sun, sending a spectacular echo of light through the 
surrounding interstellar medium \citep{bon04}. One model for this event 
suggested that we were witnessing a Very Late Thermal Pulse (VLTP), 
leading to the creation of a born-again AGB star \citep{law05}. I don't 
know if this is the correct model for V838~Mon, but in this paper I will 
discuss the born-again phenomenon, and how it relates to the evolution of 
H-deficient post-AGB stars, diffusion theory, and asteroseismology.

\section{Theoretical Background}

\subsection{Post-AGB Evolution}

As post-asymptotic-giant-branch (post-AGB) stars begin to descend the 
white dwarf cooling track, about 20\% will experience a Very Late Thermal 
Pulse \citep[VLTP;][]{ibe83}. In the process, most of the residual H in 
the envelope will be burned, but traces ($\sim10^{-11}~M_*$) are expected 
to remain \citep{her99}. The VLTP also forces the outer $\sim10^{-2}~M_*$ 
to become a uniform mixture of the remaining elements, mostly He, C, and 
O. The resulting H-deficient object becomes a born-again AGB star, and 
then continues its evolution as a hot DO white dwarf, or PG~1159 star.

% TABLE 1 %%%%%%%%%%%%%%%%%%%%%%%%%%%%%%%%%%%%%%%%%%%%%%%%%%%%%%%%%%%%%%
\begin{table*}
\centering
\begin{minipage}{100mm}
\caption{Optimal model parameters for CBS~114 and GD~358\label{tab1}}
\begin{tabular}{lcrcrcl}
\hline
Parameter                      && CBS~114 && GD~358  && Uncertainty\\
\hline
$T_{\rm eff}$~(K)$\dotfill$    && 25\,800 && 23\,100 && $\pm100$   \\
$M_*\ (M_{\odot})\dotfill$     && 0.630   && 0.630   && $\pm0.005$ \\
$\log(M_{\rm env}/M_*)\ldots$  && $-$2.42 && $-$2.92 && $\pm0.02$  \\
$\log(M_{\rm He}/M_*)\dotfill$ && $-$5.96 && $-$5.90 && $\pm0.02$  \\
$\sigma_{\rm P}$~(s)$\dotfill$ && 2.33    && 2.26    && $\cdots$   \\
\hline
\end{tabular}
\end{minipage}
\end{table*}
%%%%%%%%%%%%%%%%%%%%%%%%%%%%%%%%%%%%%%%%%%%%%%%%%%%%%%%%%%%%%%%%%%%%%%%%

\subsection{Spectral Evolution}

According to the spectral evolution theory of \citet{fw87,fw97}, DO stars 
are the progenitors of the cooler DB white dwarfs. The primary difficulty 
with this scenario is the paucity of non-DA stars with effective 
temperatures between 45,000~K and 30,000~K, the so-called ``DB gap'' 
\citep{lie86}. If there is an evolutionary connection between the hot DO 
stars and the cooler DB white dwarfs, how do we explain the missing 
H-deficient objects at intermediate temperatures? The proposed answer is 
that as a DO star cools, the small traces of H left over from the VLTP 
float to the surface through diffusion. By the time it reaches 45,000~K 
this surface H layer is thick enough to disguise the star as a DA white 
dwarf. Inside the DB gap it continues to cool as an apparent DA until a 
growing He convection zone eventually dilutes the thin surface H layer, 
and at 30,000~K the star reveals itself to be a DB. A possible relative 
overabundance of DA stars inside the DB gap supports this hypothesis 
\citep[see][their Fig.~11]{kle04}, though other problems with the theory 
may still remain \citep{pro00}.

\subsection{Diffusion Theory}

If we assume that there is an evolutionary connection between DO stars and 
DB white dwarfs, we can ask: how do the envelopes of DB stars evolve? 
Several groups have investigated this question, and they all find that 
diffusion slowly builds a pure He surface layer above the initially-uniform 
envelopes of DO stars, creating a characteristic double-layered structure 
\citep{dk95,fb02,ac04}. Such models lead to a specific prediction: at a given 
mass, {\it hotter} DB stars should have {\it thinner} surface He layers.

\section{Observations}

Fortunately, the DB stars pulsate in a range of temperatures just below 
the DB gap, allowing an asteroseismic test of this prediction. The two 
most extensively studied DB variables are GD~358 \citep{win94,vui00,kep03} 
and CBS~114 \citep{hmw02,met05}. According to the spectroscopic estimates 
of \citet{bea99} CBS~114 is up to 1500~K hotter than GD~358, so it should 
have a thinner He layer.

\section{Model-Fitting}

\cite{met05} used a parallel genetic algorithm \citep{mc03} to optimize 
the match between the observed and calculated periods ($\sigma_{\rm P}$) 
using models with 4 adjustable parameters. They searched stellar masses 
($M_*$) between 0.45 and 0.95 $M_\odot$ \citep{nap99}, effective 
temperatures ($T_{\rm eff}$) between 20,000 and 30,000~K \citep{bea99}, 
envelope masses ($M_{\rm env}$) between $10^{-2}$ and $10^{-4}~M_*$ 
\citep{dm79}, and surface He layer masses ($M_{\rm He}$) between $10^{-5}$ 
and $10^{-7}~M_*$ \citep{dk95}. For this initial experiment, they used 
pure C cores out to the 0.95 fractional mass point.

% FIGURE 1 %%%%%%%%%%%%%%%%%%%%%%%%%%%%%%%%%%%%%%%%%%%%%%%%%%%%%%%%%%%%%
\begin{figure*}[t!]
\resizebox{\hsize}{!}{\includegraphics[clip=true]{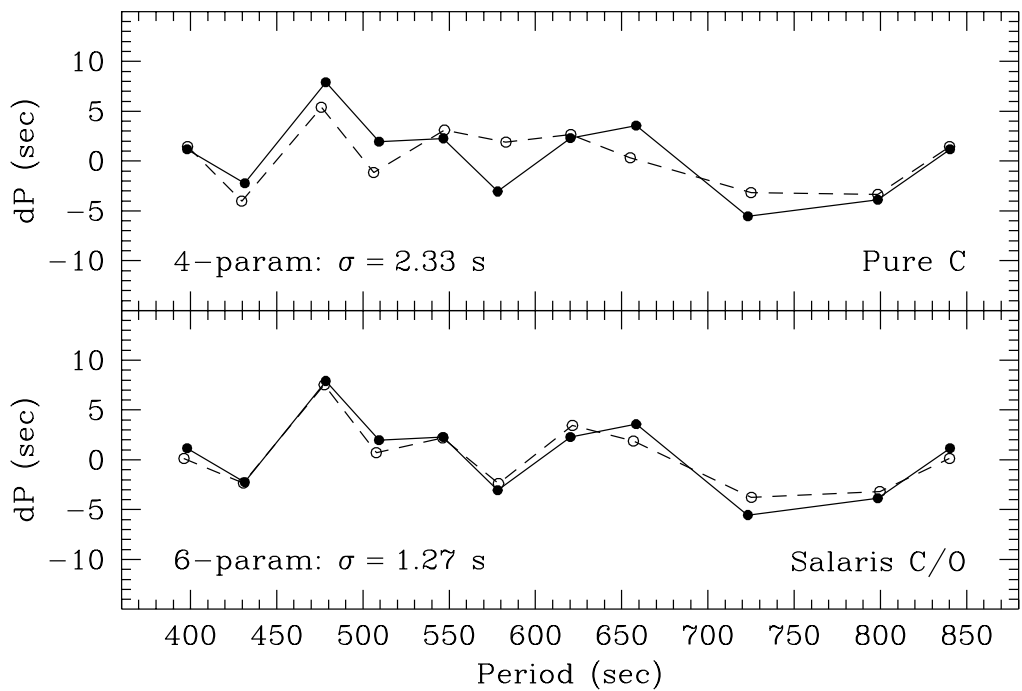}}
\caption{\footnotesize A comparison of the optimal match (open points)
to the observed periods of CBS~114 (solid points) for double-layered
envelope models with a pure C core (top panel) and an adjustable C/O 
core motivated by the calculations of \citet[][bottom panel]{sal97}.
The addition of two free parameters is expected to improve the fit
from $\sigma_{\rm P}=2.33$~s to 1.87~s, but the fit actually improves
by more than twice this amount.\label{fig1}}
\end{figure*}
%%%%%%%%%%%%%%%%%%%%%%%%%%%%%%%%%%%%%%%%%%%%%%%%%%%%%%%%%%%%%%%%%%%%%%%%

\subsection{Initial Results}

The resulting optimal model parameters for the two stars are listed in 
Table~\ref{tab1}, along with statistical uncertainties set by the 
resolution of the search. The derived mass and temperature of CBS~114 both 
agree with the spectroscopic estimates of \citet{bea99}. The mass of 
GD~358 is consistent with spectroscopy, but the derived temperature is 
about 1000~K too low. CBS~114 has a larger total envelope mass than GD~358 
and a marginally thinner surface He layer, just as predicted. Secondary 
minima in the search space hint that the models are inadequate, and that 
additional structure in the interior may be needed to explain the 
observations completely. This additional structure is most likely located 
in the core.

\subsection{Latest Results}

To explore this possibility, I added an adjustable C/O profile to the 
cores of the models. The original parameterization \citep{mwc01} fixed the 
oxygen mass fraction ($X_{\rm O}$) to its central value out to some 
fractional mass ($q$) where it then decreased linearly in mass to zero 
oxygen at $0.95~M_r/M_*$. A new version uses the same parameterization, 
but includes a physically motivated shape for the outer C/O profile, based 
on the calculations of \citet{sal97}. The optimal model for CBS~114 from 
this new version leads to a significant improvement in the fit to the 
observations (see Fig.~\ref{fig1}). \cite{met05b} presents a new series of 
model-fits using these same observations to illustrate the relative 
importance of various interior structures, and discusses the implications 
of this asteroseismic model which agrees with both diffusion theory and 
the expected nuclear burning history of the progenitor.

\section{Conclusions}

Let me summarize the main points I hope you will take away with you:
\begin{itemize}

\item Asteroseismic fits to two DBV white dwarfs are in qualitative
      agreement with diffusion theory

\item These results support a central assumption of spectral evolution
      theory, linking the DB white dwarfs to PG~1159 stars

\item Addition of a realistic C/O core profile significantly improves
      the fit to CBS~114

\item Further tests will be possible using additional DBVs from the
      Sloan Digital Sky Survey \citep[SDSS;][]{nit05}

\end{itemize}

\begin{acknowledgements}
I would like to thank my collaborators R.~E. Nather, T.~K. Watson, S.-L. 
Kim, B.-G. Park, and G. Handler for the successful dual-site observing 
campaign on CBS~114 that made these results possible. This research was 
supported by the National Science Foundation through an Astronomy \& 
Astrophysics Postdoctoral Fellowship under award AST-0401441. 
Computational resources were provided by White Dwarf Research Corporation 
(http://WhiteDwarf.org).
\end{acknowledgements}

\end{document}